\documentclass[prd,amsmath,amssymb,showpacs,superscriptaddress,nofootinbib,twocolumn]{revtex4-1}
\usepackage{graphicx}
\usepackage{epsfig,graphics,subfigure,psfrag,amsmath,amssymb}
\usepackage{dcolumn}
\usepackage{bm}
\usepackage{overpic}
\usepackage{xspace}

\usepackage{rotating}
\usepackage{color}
\usepackage{multirow}
\usepackage{colortbl}
\usepackage{epstopdf}
\usepackage[colorlinks,linkcolor=blue,anchorcolor=blue,citecolor=blue]{hyperref}

\newcommand{\jpsi}{J/\psi}
\newcommand{\etac}{\eta_{c}}

\newcommand{\pip}{\pi^+}
\newcommand{\pim}{\pi^-}
\newcommand{\piz}{\pi^0}

\newcommand{\jpsito}{J/\psi \rightarrow }

\newcommand{\etacto}{\eta_{c} \rightarrow }
\newcommand{\etapto}{\eta' \rightarrow }
\newcommand{\bfg}{\begin{figure}}
\newcommand{\efg}{\end{figure}}
\newcommand{\bitm}{\begin{itemize}}
\newcommand{\eitm}{\end{itemize}}
\newcommand{\bnum}{\begin{enumerate}}
\newcommand{\enum}{\end{enumerate}}
\newcommand{\btbl}{\begin{table}}
\newcommand{\etbl}{\end{table}}
\newcommand{\btbu}{\begin{tabular}}
\newcommand{\etbu}{\end{tabular}}

\newcommand{\beq}{\begin{equation}}
\newcommand{\edq}{\end{equation}}

\newcommand{\BESIII}{BES\uppercase\expandafter{\romannumeral3}\xspace}

\begin{document}

\title{\boldmath Search for intermediate resonances and dark gauge bosons in $J/\psi\rightarrow\gamma\pi^0\eta^{\prime}$}

\author{M.~Ablikim$^{1}$, M.~N.~Achasov$^{10,d}$, P.~Adlarson$^{59}$, S. ~Ahmed$^{15}$, M.~Albrecht$^{4}$, M.~Alekseev$^{58A,58C}$, A.~Amoroso$^{58A,58C}$, F.~F.~An$^{1}$, Q.~An$^{55,43}$, Y.~Bai$^{42}$
% ...
\\
\vspace{0.2cm}
(BESIII Collaboration)\\
\vspace{0.2cm} {\it
$^{1}$ Institute of High Energy Physics, Beijing 100049, People's Republic of China\\
% ...
$^{4}$ Bochum Ruhr-University, D-44780 Bochum, Germany\\
% ...
$^{10}$ G.I. Budker Institute of Nuclear Physics SB RAS (BINP), Novosibirsk 630090, Russia\\
% ...
$^{15}$ Helmholtz Institute Mainz, Johann-Joachim-Becher-Weg 45, D-55099 Mainz, Germany\\
% ...
$^{42}$ Southeast University, Nanjing 211100, People's Republic of China\\
$^{43}$ State Key Laboratory of Particle Detection and Electronics, Beijing 100049, Hefei 230026, People's Republic of China\\
% ...
$^{55}$ University of Science and Technology of China, Hefei 230026, People's Republic of China\\
% ...
$^{58}$ (A)University of Turin, I-10125, Turin, Italy; (B)University of Eastern Piedmont, I-15121, Alessandria, Italy; (C)INFN, I-10125, Turin, Italy\\
$^{59}$ Uppsala University, Box 516, SE-75120 Uppsala, Sweden\\
% ...
$^{d}$ Also at the Novosibirsk State University, Novosibirsk, 630090, Russia\\
% ...
}
}
\date{\today}

\begin{abstract}

  We report on an analysis of the decay $J/\psi\rightarrow\gamma\pi^0\eta^{\prime}$ using a sample of
  $(1310.6 \pm 7.0) \times ~10^6~ J/\psi$ events collected with the BESIII detector.
  We search for the CP-violating process $\eta_c\rightarrow\pi^0\eta^{\prime}$ and a dark gauge boson
  $U'$ in $J/\psi\rightarrow U^{\prime}\eta^{\prime},~U^{\prime}\rightarrow\gamma\pi^0,~\pi^0\rightarrow\gamma\gamma$.
  No evidence of an $\eta_{c}$ signal is observed in the $\pi^0\eta^{\prime}$ invariant-mass spectrum and the upper limit of the
  branching fraction is determined to be $7.2 \times~10^{-5}$ at the 90\% confidence level. We also find no evidence
  of $U'$ production and set upper limits at the 90\% confidence level on the product branching fraction
  $\mathcal{B}(J/\psi \rightarrow U^{\prime} \eta^{\prime}) \times \mathcal{B}(U^{\prime} \rightarrow \pi^{0} \gamma)$
  in the range between $(0.8 - 6.5)\times 10^{-7}$ for 0.2 $\leq m_{U^{\prime}} \leq 2.1$~GeV$/c^{2}$.
  In addition, we study the process $J/\psi\rightarrow\omega\eta^{\prime}$ with $\omega\rightarrow\gamma\pi^0$.
  The branching fraction of $J/\psi\rightarrow\omega\eta^{\prime}$ is found to be $(1.87\pm0.09\pm0.12) \times 10^{-4}$,
  where the first uncertainty is statistical and the second is systematic, with a precision that is improved by
  a factor of 1.4 over the previously published BESIII measurement.

\end{abstract}


\maketitle

\section{Introduction}

The Standard Model (SM) has been successful in explaining a wide variety of experimental data, however its
predictive power is limited by the large number of free parameters. The observation of physics beyond the SM is
needed to explain phenomena the SM cannot. Therefore, in recent years the search for new physics beyond the SM
is one of the important activities of particle physicists worldwide. The BESIII (Beijing Electron Spectrometer)
experiment is currently searching for beyond-the-SM physics using low-energy $e^+e^-$ collision data.
This is complementary to experiments conducted at the Large Hadron Collider (LHC) at CERN, which use high-energy hadron collision data.
Huge data samples accumulated by the BESIII detector and taken at center-of-mass energies corresponding to the masses of various
charmonium resonances ($J/\psi$, $\psi(3686)$ and $\psi(3770)$) offer a unique sensitivity to search for forbidden decays and dark matter particles
in the low-energy region~\cite{BESIIIyellowbook}.

Charge conjugation and parity symmetry (CP) violation has only been observed in weak interactions, which in the SM, originates from a single complex phase in the
Cabibbo-Kobayashi-Maskawa (CKM) quark-mixing matrix~\cite{CKM}. Therefore, searches for this phenomenon will provide new insights and
will help to determine whether the phase in the CKM mixing matrix is the sole source of CP violation or whether there are other sources.
The decay of an $\eta_c$ ($J^{PC}=0^{-+}$) to two pseudoscalar mesons is forbidden due to CP conservation. The observation of these forbidden
decays will be a clear indication of new physics beyond the SM.
Using a sample of 225 million $\jpsi$ events, BESIII report the results of the search for $\etacto\pip\pim$ and $\etacto\piz\piz$ and
upper limits on the branching fractions are presented at the 90\% confidence level (C.L.)~\cite{etac}.
In this paper, we present the first experimental search for $\etacto\piz\eta'$.

Except for gravitational effects, we still know very little about the constituents and interactions of dark matter. One possible model candidate
for dark matter is an additional gauge boson~\cite{Bboson,Bboson1}. If this additional boson corresponds to an extra $U(1)$ gauge symmetry,
it is referred to as a ``dark photon''. A dark photon with a mass in the range from MeV/$c^2$ to GeV/$c^2$ can be used to explain the feature of
recently observed astrophysical anomalies~\cite{anolomy} as well as a 3-4$\sigma$ deviation in the muon anomalous magnetic moment
between the measurement and the SM prediction~\cite{deviation}.
This new gauge boson, referred to as $U'$,
has the same quantum number, $J^{PC}$ = $1^{--}$, as the $\omega$ meson. In the past, BESIII has reported on a search for the dark gauge photon ($\gamma^{\prime}$)
in the initial-state radiation (ISR) reactions $e^+e^- \rightarrow \gamma^{\prime} \gamma_{\rm ISR} \rightarrow l^+l^-\gamma_{\rm ISR}$ $(l= \mu, e)$
~\cite{besiii2017} and electromagnetic Dalitz decays $J/\psi\rightarrow\gamma^{\prime}\eta/\eta^{\prime} \rightarrow e^+e^-\eta/\eta^{\prime}$~\cite{gammaeta,gammaeta'}.
The same ISR method has been used by the BaBar
experiment~\cite{babar2014,babar2017}. The BELLE and KLOE
collaboration report a search for a dark vector gauge boson decaying to $\pi^{+}\pi^{-}$, where the dark vector gauge
boson mass spans a range from 290 to 520~MeV/$c^2$~\cite{belle} and 527 to 987~MeV/$c^2$~\cite{kloe}, respectively.

In this paper, using a sample of 1.31 $\times~10^9~\jpsi$ events collected with the BESIII detector, we present the first study
of $J/\psi\rightarrow\gamma\pi^0\eta^\prime$, which allows us to search for the CP-violating decay of $\eta_c\rightarrow\pi^0\eta^\prime$
and to search for a new gauge boson~\cite{Bboson} by investigating the $\gamma\pi^0$-mass spectrum. Additionally, we present
the most accurate measurement of the $J/\psi\rightarrow\omega\eta'$ branching fraction
(current BESIII measurement value is (2.08 $\pm$ 0.30 $\pm$ 0.14) $\times$ $10^{-4}$~\cite{omegaetap}).

\section{THE BESIII EXPERIMENT AND MONTE CARLO SIMULATION}

The BESIII detector is a cylindrical magnetic
spectrometer~\cite{Ablikim:2009aa} located at the Beijing Electron
Positron Collider (BEPCII)~\cite{Yu:IPAC2016-TUYA01}, with an acceptance of
charged particles and photons of 93\% over $4\pi$ solid angle. The BESIII
detector consists of a helium-based
multilayer drift chamber (MDC), a plastic scintillator time-of-flight
system (TOF), and a CsI~(Tl) electromagnetic calorimeter (EMC),
which are all enclosed in a superconducting solenoidal magnet
providing a 1.0~T (0.9~T in
2012) magnetic field. The solenoid is supported by an
octagonal flux-return yoke with resistive plate counter muon
identifier modules interleaved with steel. The
charged-particle momentum resolution at $1~{\rm GeV}/c$ is
$0.5\%$, and the $dE/dx$ resolution is $6\%$ for the electrons
from Bhabha scattering. The EMC measures photon energies with a
resolution of $2.5\%$ ($5\%$) at $1$~GeV in the barrel (end cap)
region. The time resolution of the TOF barrel part is 68~ps, while
that of the end cap part is 110~ps.

Simulated samples produced with the {\sc
geant4}-based~\cite{geant4} Monte Carlo (MC) package which
includes the geometric description~\cite{BesGDML,GDMLMethod} of the BESIII detector and the
detector response, are used to determine the detection efficiency
and to estimate the backgrounds. The simulation includes the beam
energy spread and ISR in the $e^+e^-$
annihilations generated using the {\sc kkmc} package~\cite{ref:kkmc}.
The inclusive MC sample consists of the production of the $J/\psi$
resonance, and the continuum processes incorporated in {\sc
kkmc}~\cite{ref:kkmc}.
The known decay modes are generated using the {\sc
evtgen} package~\cite{ref:evtgen} using branching fractions taken from the
Particle Data Group (PDG)~\cite{pdg}, and the remaining unknown decays
from the charmonium states with the {\sc
lundcharm} package~\cite{ref:lundcharm}. The final-state radiations (FSR)
from charged final-state particles are incorporated with the {\sc
photos} package~\cite{photos}.

The three-body decay of $\jpsito\gamma\piz\eta'$ without any intermediate states is
simulated with a model based on a phase-space distribution of the final-state particles.
The decays of $\jpsito\gamma\etac,~U'\eta',~\gamma\eta'$ and $\omega\eta'$ are generated with an angular
distribution of $1+\cos^2\theta_{\gamma}$, where $\theta_{\gamma}$ is the angle of radiative photon
relative to the positron beam direction in the $\jpsi$-rest frame, while the subsequent $\etac~(\eta')$
decays are generated with a phase-space model and the $U'~(\omega)\rightarrow\gamma\piz$ decay is modeled
by a $P$-wave~\cite{ref:evtgen}.

\section{Event Selection}

Candidates of $\jpsito\gamma\piz\eta',~\etapto\pip\pim\eta,~\piz\rightarrow\gamma\gamma,~\eta\rightarrow\gamma\gamma$
are required to have two oppositely-charged tracks with a zero net charge and at least five photon candidates.
All charged tracks must originate from the interaction point with a distance of closest approach less than 10~cm in the beam direction
and less than 1~cm in the transverse plane. Their polar angles, $\theta$, with respect to the beam direction are required to
satisfy $|\cos\theta|<$~0.93. Particle identification (PID) for charged pions is performed by exploiting the TOF information and
the specific ionization energy loss, $dE/dx$, measured by the MDC.
The TOF and $dE/dx$ information are combined to form PID probability for the pion, kaon, and proton hypotheses; each track is
assigned to the particle type that corresponds to the hypothesis with the highest probability.

Electromagnetic showers are reconstructed from clusters of firing EMC crystals.
The energy deposited in nearby TOF counters is included to improve the reconstruction efficiency and energy resolution.
The showers of the photon candidate must have a minimum energy of 25~MeV in the barrel region ($|\cos\theta|<0.80$) and
50~MeV in the end-cap region (0.86~$<|\cos\theta|<$~0.92). To suppress showers originating from charged particles,
a photon candidate must be separated by at least $10^\circ$ from the nearest charged track.
An EMC shower timing requirement, $0\leq t \leq 700~\rm ns$, is applied to suppress noise and energy deposits unrelated to the event.

After selecting the charged tracks and showers, a four-constraint (4C) kinematic fit to the $\jpsito\pip\pim5\gamma$ hypothesis
is performed using energy-momentum conservation. For events with more than five photon candidates, the combination with the
smallest $\chi^2_{4C}$ is retained.
To suppress background events with six photons in the final states, the $\chi^2_{4C}$ of the $\pip\pim5\gamma$
hypothesis is required to be less than that for the $\pip\pim6\gamma$ hypothesis.

To distinguish the photon from $\piz$ and $\eta$ decays, we define the variable
$\chi^2_{\piz\eta} \equiv (\frac {M_{\gamma\gamma}-m_{\piz}}{\sigma_{\pi^0}})^2 + (\frac {M_{\gamma\gamma}-m_{\eta}}{\sigma_\eta})^2$.
This variable is used to choose from the five photon candidates two pairs of photons with two-photon invariant masses ($M_{\gamma\gamma}$)
closest to the nominal $\pi^0$ ($m_{\piz}$) and $\eta$ ($m_{\eta}$) masses.
$\sigma_{\pi^0}$ ($\sigma_{\eta}$) refers to the experimental mass resolution for a $\piz$ ($\eta$) decay. The four-photon combination with the smallest value for $\chi^2_{\piz\eta}$ is chosen.

To improve the mass resolution and to further suppress background events, we subsequently perform a five-constraint kinematic (5C)
fit imposing energy-momentum conservation and a $\eta$-mass constraint under the hypothesis of $\pip\pim\gamma\gamma\gamma\eta$,
where the $\eta$ candidate is reconstructed with the selected pair of photons as described above. Events with a $\chi^2_{5C}$ less than 30 are
accepted for further analysis.

To select $\piz$ candidates, the invariant mass of the two photons from $\piz$ decay, $M_{\gamma\gamma}$, must satisfy $|M_{\gamma\gamma}-m_{\piz}|<$~15~MeV/$c^2$.
To suppress background events with multi-$\piz$ in the final states, we require that the invariant mass of the radiative photon and
one photon from the $\eta$ decay is outside the $\pi^0$-mass region of [0.115, 0.155]~GeV/$c^2$.
To select $\eta'$ candidates, we calculate for each event the $\pip\pim\eta$ invariant mass, $M_{\pip\pim\eta}$, and require that
$|M_{\pip\pim\eta}-m_{\eta'}|<$~15~MeV/$c^2$, where $m_{\eta'}$ is the nominal $\eta'$ mass.

\section{\boldmath Search for $\etacto\piz\eta'$}

\begin{figure*}
  \centering
  \vskip -0.2cm
  \hskip -0.4cm \mbox{
  \begin{overpic}[width=0.8\textwidth]{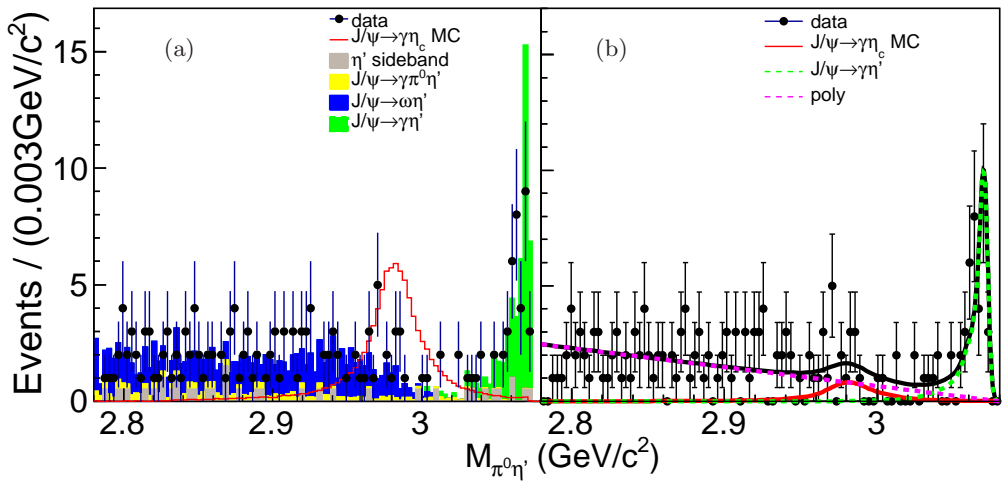}
  \put(15,40){(a)}
  \put(55,40){(b)}
  \end{overpic}
  }
  \vskip -0.7cm
  \hskip 0.5cm
  \caption{The $\pi^{0} \eta^{\prime}$-mass spectrum. The black dots with error bars are data. (a) The histogram with the red line represents
    the extracted lineshape of the signal process $J/\psi \rightarrow \gamma \eta_{c}$. The yellow area shows the MC distribution of
    $J/\psi \rightarrow \gamma \pi^{0} \eta^{\prime}$, the green area corresponds to the MC distribution of $J/\psi \rightarrow \gamma \eta^{\prime}$,
    the blue area shows the MC distribution of $J/\psi \rightarrow \omega \eta^{\prime}$, and the gray area represents the non-$\eta^{\prime}$
    contributions obtained from $\eta^{\prime}$-sideband data.
    (b) The black line shows the result of a maximum-likelihood fit. The red histogram shows the contribution of the $\eta_{c}$ signal,
    the green dashed line represents the $J/\psi \rightarrow \gamma \eta^{\prime}(\eta^{\prime} \rightarrow \eta \pi^{+} \pi^{-}, \eta \rightarrow \gamma \gamma)$
    background contribution, and the pink dashed line depicts other non-peaking background contributions described by a first-order Chebychev polynomial.
  }\label{pi0etap}
\end{figure*}

After applying the selection criteria, we obtain the $\piz\eta'$ invariant-mass distribution as shown in Fig.~\ref{pi0etap}. No evident $\eta_c$ peak is seen.
Using a MC sample of inclusive $1.2\times 10^9$ $J/\psi$ decays, we found that the dominant background events are from decays with the $\eta'$ as an
intermediate state, such as  $\jpsito\gamma\piz\eta'$, $\jpsito\omega\eta^\prime$ and $\jpsito\gamma\eta'$, and the corresponding contributions
are displayed in Fig.~\ref{pi0etap}(a) as well. Other background contributions (non-$\eta'$ background) are estimated from events for which the
reconstructed $\eta'$ mass falls within the $\eta'$-sideband regions (0.903~$<M_{\pip\pim\eta}<$~0.933~GeV/$c^{2}$ and 0.983~$<M_{\pip\pim\eta}<$~1.013~GeV/$c^{2}$).
The sum of the above contributions gives a reasonable description of data.

The statistical significance of finding the $\etac$ signal is calculated to be 1.7$\sigma$ using
$\sqrt{-2{\rm ln}({\mathcal L}^{\rm stat}_0/{\mathcal L}^{\rm stat}_{\rm max})}$, where ${\mathcal L}^{\rm stat}_{\rm max}$ and ${\mathcal L}^{\rm stat}_0$
are the maximum-likelihood values with the signal yield left free and fixed at zero, respectively. Since no evident $\etac$ signal is seen in $M_{\piz\eta'}$,
a Bayesian method is used to obtain the upper limit of the signal yield at the 90\% C.L.. To determine the upper limit on the $\eta_{c}$ signal, a series
of unbinned maximum-likelihood fits are performed to the $\piz\eta'$-mass spectrum with a varying number of expected $\eta_{c}$ signals.
From this, we obtain the dependence of the likelihood on the number of signal events from which we extract the upper limit.
Figure~\ref{pi0etap}(b) depicts the result of one of these fits corresponding to the case with a maximum likelihood and a signal yield of
$N_{\rm sig}=13.2 \pm 8.2$.
In the fit, the probability density function (PDF) of the signal $\eta_{c}$ response and the background shape from
the $\jpsito\gamma\eta'$ channel are extracted from MC simulations. The absolute yield of this background
is fixed in the fit according to the published branching fractions~\cite{pdg}.
The other non-peaking background is described by a first-order Chebychev polynomial which parameters
are allowed to vary freely.

The systematic uncertainties that affect the upper limits on the branching fraction of $\eta_{c}\rightarrow\pi^0\eta^{\prime}$ are considered in two categories: additive and multiplicative. The additive systematic uncertainties on the fit range and background shapes are considered by varying the fit range and changing the background shape. The maximum upper limit among these cases is adopted and the corresponding distribution is illustrated in Fig.~\ref{pi0etap}(b). All of the systematic uncertainties, which are listed in Table~\ref{systematic}, excluding the fit range and background shape, are considered as the multiplicative systematic uncertainties. The effects of multiplicative systematic uncertainties are introduced in section RESULTS.

\section{\boldmath Search for dark photon in $U'\rightarrow\gamma\piz$ decay}

\begin{figure}
  \centering
  \vskip -0.1cm
  \hskip -0.4cm \mbox{
  \begin{overpic}[width=0.45\textwidth]{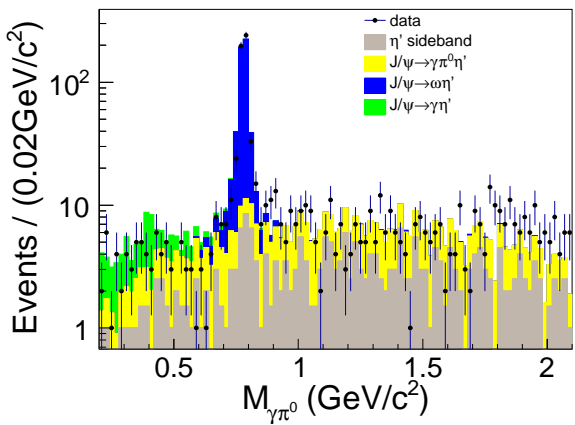}
  \end{overpic}
  }
  \vskip -0.7cm
  \hskip 0.5cm
  \caption{The $\gamma \pi^{0}$ invariant-mass spectrum. The black dots with error bars are data. The various shaded histograms
    are described in the caption of Fig.~\ref{pi0etap}.
  }\label{gammapi0}
\end{figure}

Using the same selection criteria as used to search for $\eta_c\rightarrow \pi^0 \eta^\prime$, we study the $\gamma\piz$-mass ($M_{\gamma\piz}$) distribution
as shown in Fig.~\ref{gammapi0}. A clear $\omega$ peak from $\jpsito\omega\eta'$ decays can be observed. There is also a small background contribution
from $\jpsito\gamma\eta'$ decays which is smoothly distributed in the low-mass region of the $M_{\gamma\piz}$ distribution.
The contributions from non-$\eta'$ backgrounds are described by events that are selected in the $\eta'$-sideband regions,
0.903~$<M_{\pip\pim\eta}<$~0.933~GeV/$c^{2}$ and 0.983~$<M_{\pip\pim\eta}<$~1.013~GeV/$c^{2}$.

\begin{figure}
  \centering
  \vskip -0.2cm
  \hskip -0.4cm \mbox{
  \begin{overpic}[width=0.45\textwidth]{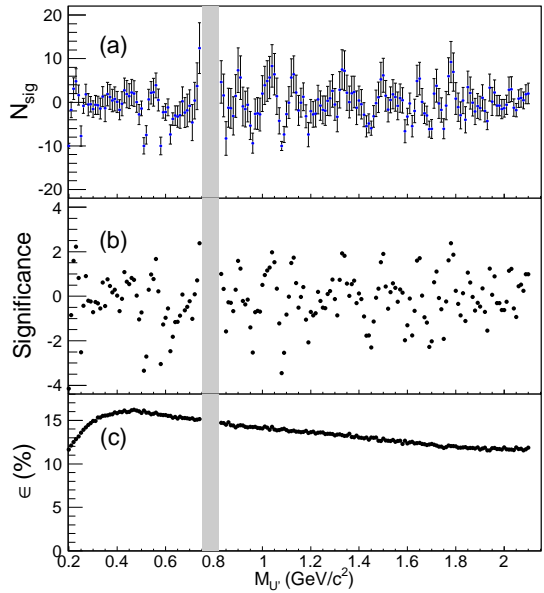}
  \end{overpic}
  }
  \vskip -0.5cm
  \hskip 0.5cm
  \caption{(a) The number of extracted signal events, (b) statistical signal significance, and (c) the detection efficiency as a function of
    $M_{U'}$ in the range of 0.2~$\leq M_{U'} \leq$~2.1~GeV/$c^2$. The region of the $\omega$ resonance is indicated by the gray band and
    excluded from the $U'$ search.
  }\label{Usig}
\end{figure}

We search for the $U'$ signal in step of 10~MeV/$c^2$ in the $M_{\gamma\piz}$ distribution ranging from 0.2~GeV/$c^2$ to 2.1~GeV/$c^2$
and excluding the mass region around the $\omega$ peak (0.75 to 0.82~GeV/$c^2$). The mass resolution of a $U^{\prime}$ signal
has been evaluated using signal MC events generated at 183 different $U^{\prime}$-mass ($M_{U'}$) hypotheses points with a negligible width.
Depending upon the $U^\prime$ mass, the resolutions vary in the range between 3.6 and 10.4~MeV/$c^2$. We perform a series of unbinned extended
maximum-likelihood fits to the $M_{\gamma\piz}$ distribution to determine the number of signal candidates as a function of $M_{U'}$ in the interval of
0.2~$\leq M_{U'} \leq$~2.1~GeV/$c^2$. In the fit, the $U'$ signal and the tail of the $\omega$ signal are described by MC-simulated shapes,
and the remaining background contribution is modeled with a linear Chebychev polynomial. To take into account the additive systematic uncertainties related to the fits,
alternative fits with different fit range and background shape also performed, and the maximum upper limit among these cases has been selected.
The number of extracted signal events, the significance, and the detection efficiency as a function of $M_{U'}$ are shown in Fig.~\ref{Usig}.
The largest local significance defined as before is computed to be 2.4$\sigma$ at $M_{U'}$~=~1.78~GeV/$c^2$. No significant signal for
$U'\rightarrow\gamma\piz$ is found.

\section{\boldmath Branching fraction measurement of $\jpsito\omega\eta'$}

\begin{figure}
  \centering
  \vskip -0.2cm
  \hskip -0.4cm \mbox{
  \begin{overpic}[width=0.45\textwidth]{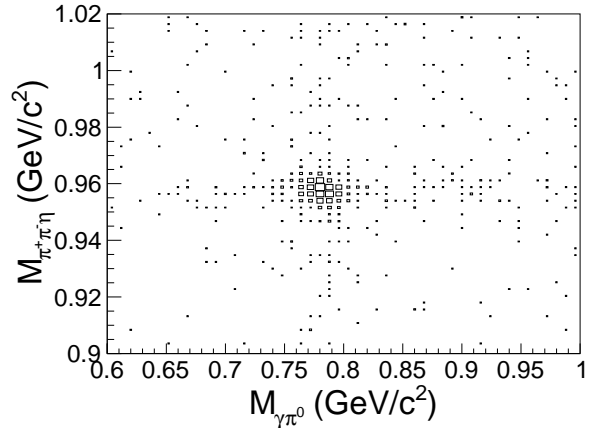}
  \end{overpic}
  }
  \vskip -0.5cm
  \hskip 0.5cm
  \caption{A two-dimensional distribution depicting the relation between the reconstructed $\pi^{+}\pi^{-}\eta$ and $\gamma\pi^0$ masses.
    The size of each box scales with the number of events found in that particular bin.
  }\label{Scatter}
\end{figure}

Figure~\ref{Scatter} shows the mass distribution of $M_{\pip\pim\eta}$ versus $M_{\gamma\piz}$.
Events originating from the $\jpsito\omega\eta'$ decay are clearly visible. To extract the number of $\omega\eta'$ events,
an unbinned extended maximum-likelihood fit using a two-dimensional (2-D) PDF including both variables, $M_{\pip\pim\eta}$ and $M_{\gamma\piz}$,
with the requirements of 0.6~$< ~M_{\gamma\piz} <$~1.0 GeV/$c^2$ and 0.908~$<~M_{\pip\pim\eta}~<$~1.008 GeV/$c^2$. Assuming
zero correlation between the two discriminating variables $M_{\gamma\piz}$ and $M_{\pip\pim\eta}$, the composite PDF in the 2-D
fit is constructed as follows
$$\begin{aligned}
F&=N_{\rm sig}\times(F^{\omega}_{\rm sig}\cdot F^{\eta'}_{\rm sig}) \\
&+ N^{\rm non-\omega}_{\rm bkg}\times(F^{\eta'}_{\rm sig}\cdot F^{\rm non-\omega}_{\rm bkg}) \\
&+ N^{\rm non-\eta'}_{\rm bkg}\times(F^{\omega}_{\rm sig}\cdot F^{\rm non-\eta'}_{\rm bkg})\\
&+ N^{\rm non-\omega\eta'}_{\rm bkg}\times(F^{\rm non-\omega}_{\rm bkg}\cdot F^{\rm non-\eta'}_{\rm bkg}),
\end{aligned}$$
where the signal shapes for the $\omega$ ($F^{\omega}_{\rm sig}$) and $\eta'$ ($F^{\eta'}_{\rm sig}$) responses are modeled
with a relativistic Breit-Wigner (BW) function convoluted with a Gaussian function. The widths and masses of the $\omega$ and $\eta'$
are fixed in the fit. The parameters of the Gaussian function are free in the fit.
$N_{\rm sig}$ is the number of $\jpsito\omega\eta',\omega\rightarrow\gamma\piz, \eta'\rightarrow\pip\pim\eta$ signal events.
The backgrounds are divided into three categories, namely non-$\omega$ peaking background, non-$\eta'$ peaking background,
and non-$\omega\eta'$ background. The parameters $N^{\rm non-\omega}_{\rm bkg}$,
$N^{\rm non-\eta'}_{\rm bkg}$, and $N^{\rm non-\omega\eta'}_{\rm bkg}$ are the corresponding three background yields.
The background shapes, $F^{\rm non-\omega}_{\rm bkg}$ and $F^{\rm non-\eta'}_{\rm bkg}$,
related to $M_{\gamma\piz}$ and $M_{\pip\pim\eta}$, respectively, are described by first-order Chebychev polynomials
and all their corresponding parameters are free in the fit.

\begin{figure*}
  \includegraphics[width=6.0in]{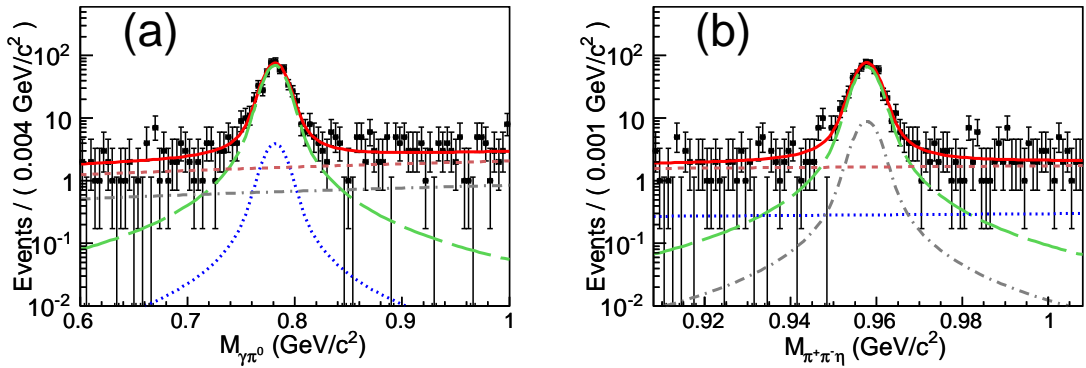}
  \vskip -0.2cm
\caption{Projection plots of (a) $M_{\gamma\piz}$ and (b) $M_{\pip\pim\eta}$ distributions in the decay chain
  of $\jpsito\omega\eta', \omega\rightarrow\gamma\piz, \eta'\rightarrow\pip\pim\eta$. The dots with error bars correspond to data,
  the solid curve shows the result of the fit including both signal and background distributions. The long-dashed curve corresponds
  to the contribution of the $\omega\eta'$ signal, the dotted curve shows the contribution of the non-$\eta'$ peaking background,
  the dot-dashed curve shows the contribution of the non-$\omega$ peaking background, and the short-dashed curve represents
  the non-$\omega\eta'$ background part.} \label{omegaetap}
\end{figure*}

The best fit results to $N_{\rm sig} = 506\pm25$ signal events. The projection plots of the fit on the $M_{\gamma\piz}$
and $M_{\pip\pim\eta}$ distributions are shown in Fig.~\ref{omegaetap} (a) and Fig.~\ref{omegaetap} (b), respectively.

\section{systematic uncertainty}

The sources of systematic uncertainties and their corresponding contributions to the measurements of the upper limits and branching fraction are
summarized in Table~\ref{systematic}. Below, we briefly describe the procedure that has been applied to obtain the various systematic uncertainties.

The uncertainty of the number of $\jpsi$ events is determined to be 0.54\% by an analysis of inclusive hadronic events in $\jpsi$ decays~\cite{jpsi}.

The uncertainty of the MDC tracking efficiency for each charged pion is studied by analyzing a nearly-background free sample of $\jpsito\rho\pi$ events.
The difference between data and MC simulation is less than 1.0\% for each charged track~\cite{MDC} which value is taken as a systematic uncertainty.
Similarly, the uncertainty related to the PID efficiencies of pions
is also studied with the data sample, $\jpsito\rho\pi$, and the average difference of the PID efficiencies between data and MC simulation
is determined to be 1.0\% for each charged pion, which is then taken as the corresponding systematic uncertainty.
The photon detection efficiency is studied with the control sample $\jpsito\pip\pim\piz$~\cite{photon}. The difference in efficiency between data and
predicted by MC simulations is found to be 0.5\% per photon in the EMC barrel and 1.5\% per photon in the end-cap part of the EMC.
In our case, the uncertainty is on average 0.6\% per photon which value is obtained by weighting the uncertainties according to the angular
distribution of the five photons found in our data sample. Thus, the uncertainty associated with the five reconstructed photons is 3.0\%.

The uncertainty associated with the 5C kinematic fits comes from the inconsistency of the track helix parameters between data and MC simulation.
The helix parameters for the charged tracks of MC samples are corrected to eliminate part of the inconsistency, as described in Ref.~\cite{helix}.
We take half of the differences on the selection efficiencies with and without the correction as an estimate of the corresponding
systematic uncertainties, which results in 0.4\%.

Due to the difference in the mass resolution between data and MC, the uncertainty related to the $\eta'$ and $\piz$ mass-window requirements
are investigated by smearing the MC simulation in accordance with the signal shape of data. The difference of the detection efficiency before and
after smearing are assigned as the systematic uncertainty for the $\eta'$ and $\piz$ mass-window requirements and found to be 0.2\% and 1.1\%, respectively.

The systematic uncertainty related to the finite statistics used by the MC simulation to obtain the overall reconstruction efficiency is
calculated as $\sqrt{\frac {\epsilon(1-\epsilon)} n}$, where $\epsilon$ is the detection efficiency and $n$ is the number of generated MC events of the signal process.
The corresponding systematic uncertainty is determined to be 1.0\%.

The systematic uncertainties related to the choice of fit range and background shapes in the
$\eta_{c}\rightarrow\pi^0\eta^{\prime}$ and $U^{\prime}\rightarrow\gamma\pi^0$ searches
are already accounted for in the analysis procedure that is applied to obtain the maximum upper limit of the signal yield.
Here we, therefore, only consider these uncertainties for the $J/\psi\rightarrow\omega\eta^{\prime}$ study.
To study the uncertainty from the fit range, the fit is repeated with different fit ranges, and the resultant largest difference in the signal yield, 1.8\%,
is taken as the systematic uncertainty. The uncertainty associated with the background shape in the fits to the $M_{\gamma\piz}$ distribution
is estimated using alternative fits by changing the linear Chebychev polynomial to a second-order Chebychev polynomial.
The difference in signal yield (0.6\%) is taken as the systematic uncertainty.

The uncertainty associated with the 2-D fits of the $J/\psi\rightarrow\omega\eta^{\prime}$ channel is estimated by taking the mass and width of the BW function
as free parameters in the fit. The change in signal yield (1.0\%) is taken as the systematic uncertainty. The systematic uncertainty due to the $\piz$ veto is evaluated
by varying the requirement on the mass window, and the difference in yield compared to the nominal choice (1.1\%) is assigned as the systematic uncertainty.

The branching fractions of the intermediate processes of
$\jpsito \gamma \eta_{c}$, $\omega \rightarrow \gamma \piz$, $\eta^{\prime} \rightarrow \pip\pim\eta$, $\eta \rightarrow \gamma \gamma$
and $\piz \rightarrow \gamma \gamma$ are taken from the PDG~\cite{pdg} and their errors are considered as a source of systematic uncertainty.

For each case, the total systematic uncertainty is given by the quadratic sum of the individual contributions, assuming all sources to be independent.

\begin{table}
    \begin{center}
    \caption {The systematic uncertainties and their sources for the (product) branching fractions of the two upper-limit studies ($\eta_c$ and $U^\prime$) and
          of the $J/\psi\rightarrow\omega\eta^{\prime}$ channel. All values are in given in percentage.}
    \label{systematic}
    \begin{tabular}{lccc}\hline\hline
                Source                                                &~~$\eta_{c}$     &~~ $U^{\prime}$ &~~     $J/\psi\rightarrow\omega\eta^{\prime}$     \\ \hline
      Number of $J/\psi$ events                                       &~~ $0.54$        &~~ $0.54$               &~~ $0.54$         \\
      MDC Tracking                                                    &~~ $2.0$         &~~ $2.0$                &~~ $2.0$        \\
      Particle identification                                         &~~ $2.0$         &~~ $2.0$                &~~ $2.0$        \\
      Photon reconstruction                                           &~~ $3.0$         &~~ $3.0$                &~~ $3.0$        \\
      5C kinematic fit                                                &~~ $0.4$         &~~ $0.4$                &~~ $0.4$        \\
      $\eta^{\prime}$ mass window                                     &~~ $0.2$         &~~ $0.2$                &~~ $0.2$        \\
      $\pi^{0}$ mass window                                           &~~ $1.1$         &~~ $1.1$                &~~ $1.1$        \\
      MC efficiency                                                   &~~ $1.0$         &~~ $1.0$                &~~ $1.0$        \\
      Fit range                                                       &~~ $-$           &~~ $-$                  &~~ $1.8$     \\
      Background shape                                                &~~ $-$           &~~ $-$                  &~~ $0.6$      \\
      2-D fit                                                         &~~ $-$           &~~ $-$                  &~~ $1.0$     \\
      $\pi^{0}$ veto                                                  &~~ $1.1$         &~~ $1.1$                &~~ $1.1$        \\
      $\mathcal{B}$($J/\psi \rightarrow \gamma \eta_{c}$)             &~~ $23.5$        &~~ $-$                  &~~ $-$      \\
      $\mathcal{B}$($\omega \rightarrow \gamma \pi^0$)                &~~ $-$           &~~ $-$                  &~~ $3.4$      \\
      $\mathcal{B}$($\eta^{\prime} \rightarrow \pi^{+} \pi^{-} \eta$) &~~ $1.6$         &~~ $1.6$                &~~ $1.6$        \\
      $\mathcal{B}$($\eta \rightarrow \gamma \gamma$)                 &~~ $0.5$         &~~ $0.5$                &~~ $0.5$        \\
      $\mathcal{B}$($\pi^{0} \rightarrow \gamma \gamma$)              &~~ $0.03$        &~~ $0.03$               &~~ $0.03$         \\ \hline
      Total                                                           &~~ $24.0$        &~~ $4.9$                &~~ $6.3$        \\ \hline\hline
    \end{tabular}
    \end{center}
\end{table}

\section{Results}

We observe no evidence for $\etacto\piz\eta'$, nor for $U'\rightarrow\gamma\piz$ decays. The upper limits of the branching fraction of $\eta_c\rightarrow\pi^0\eta^{\prime}$ are estimated by a likelihood scan method, which takes into account the multiplicative systematic uncertainties as follows
$${\mathcal L} (\mathcal{B})=\int_{-1}^{1} {\mathcal L}^{\rm stat}(\mathcal{B}^{\prime})e^{-\frac{\Delta^2}{2\sigma_{\rm syst}^{2}}}\, d\Delta .$$
Here, $\mathcal{B}^{\prime}=(1+\Delta)\mathcal{B}$, where $\Delta$ is the relative deviation of the estimated branching fraction from the nominal value, and $\sigma_{\rm syst}$ is the multiplicative systematic uncertainties which are proportional to the assumed branching fraction. The proportional constant is the total systematic uncertainty given in Table~\ref{systematic}.

The branching fraction for a particular decay process is computed as
$$\mathcal{B}(X \to Y)=\frac{N_{\rm sig}}{\epsilon\times \mathcal{B}},$$
\noindent where $N_{\rm sig}$ is the number of extracted signal yield, $\epsilon$ is the signal selection efficiency,
and $\mathcal{B}$ is the secondary branching fraction of the corresponding decay process.

\begin{figure}
  \centering
  \vskip -0.2cm
  \hskip -0.4cm \mbox{
  \begin{overpic}[width=0.45\textwidth]{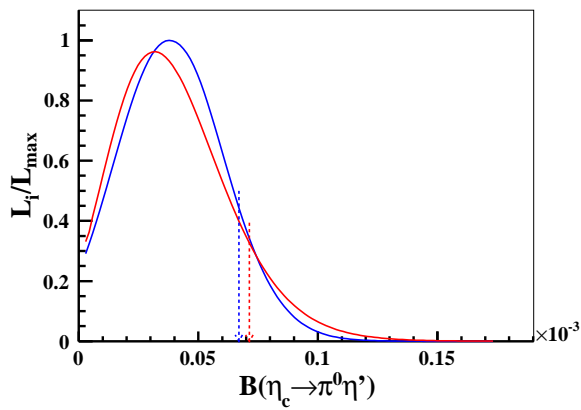}
  \end{overpic}
  }
  \vskip -0.5cm
  \hskip 0.5cm
  \caption{The distribution of the normalized likelihood scan for $J/\psi\rightarrow\gamma\eta_c(\eta_c\rightarrow\pi^0\eta^{\prime})$ candidates.
    The blue and red curves describe the smoothed likelihood curves before and after the inclusion of the multiplicative systematic uncertainty.
    The blue and red arrows show the upper limit on the signal yield at 90\% C.L..
  }\label{likelihood}
\end{figure}

The normalized likelihood distribution for $J/\psi\rightarrow\gamma\eta_c(\eta_c\rightarrow\pi^0\eta^{\prime})$ candidates is shown in Fig.~\ref{likelihood}.
The upper limit at the 90\% C.L. of the signal yield ($N_{\rm UL}$) and detection efficiency are determined to be 24.5 and 9.3\% respectively,
resulting in a branching fraction
$\mathcal{B}(\etac\rightarrow\pi^0\eta^{\prime})$ of less than $7.2\times10^{-5}$.

\begin{figure}
  \centering
  \vskip -0.2cm
  \hskip -0.4cm \mbox{
  \begin{overpic}[width=0.45\textwidth]{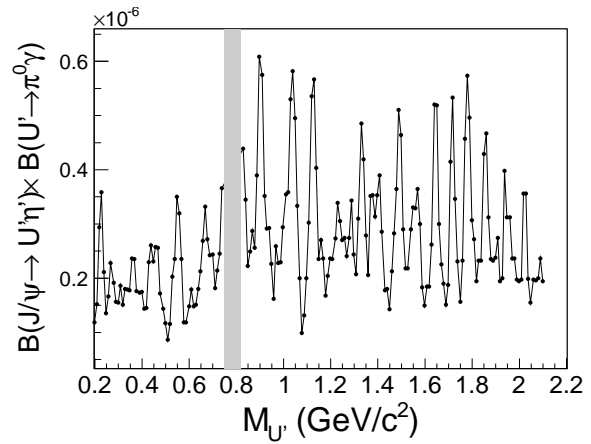}
  \end{overpic}
  }
  \vskip -0.5cm
  \hskip 0.5cm
  \caption{(a) The upper limit at the 90\% C.L. on the product branching fraction $\mathcal{B}(J/\psi \rightarrow U^{\prime} \eta') \times \mathcal{B}(U^{\prime} \rightarrow \pi^{0} \gamma)$. The region of the $\omega$ resonance indicated by the gray band is excluded from the $U'$ search.
  }\label{Ubr}
\end{figure}

We compute the upper limit on the product branching fraction $\mathcal{B}(J/\psi \rightarrow U^{\prime} \eta') \times \mathcal{B}(U^{\prime} \rightarrow \pi^{0} \gamma)$
at the 90\% C.L. as a function of $M_{U'}$ using a Bayesian method after incorporating the systematic uncertainty by smearing the likelihood
curve with a Gaussian function with a width of the systematic uncertainty.
As shown in Fig.~\ref{Ubr}, the combined limits on product branching fraction $\mathcal{B}(J/\psi \rightarrow U^{\prime} \eta') \times \mathcal{B}(U^{\prime} \rightarrow \pi^{0} \gamma)$ are established at the level of $(0.8- 6.5)\times10^{-7}$ for 0.2~$\leq M_{U^{\prime}} \leq 2.1$~GeV$/c^{2}$.

With a detection efficiency of 14.9\%, obtained from a MC simulation, we obtain a branching fraction for the $\jpsito\omega\eta'$ process of
$(1.87 \pm 0.09 \pm 0.12) \times 10^{-4}$, where the first uncertainty is statistical and the second systematic.

\section{Summary}

Using a sample of $(1310.6 \pm 7.0) \times 10^6~ \jpsi$ events collected with the BESIII detector,
  the decay of $\jpsito\gamma\eta'\piz$ is studied.
 We search for the CP-violating decay $\etacto\piz\eta'$ and a dark gauge boson $U'$ in $\jpsito U'\eta',~U'\rightarrow\gamma\piz,~\piz\rightarrow\gamma\gamma$.
 No significant $\eta_{c}$ signal is observed in the $\piz\eta'$ invariant-mass spectrum, and the upper limit on the branching fraction is determined
 to be $7.2\times10^{-5}$ at the 90\% C.L.. Except for a clear $\omega$ peak in the $\gamma\pi^0$ mass spectrum, no significant excess is seen for
 any mass hypothesis in the range of $0.2 \leq M_{U^{\prime}} \leq 2.1$ GeV$/c^{2}$.
 The upper limits on the product branching fractions are calculated to be (0.8 - 6.5)$\times10^{-7}$ at the 90\% C.L..

 In addition, the branching fraction of $\jpsito\omega\eta'$ is measured to be $(1.87\pm0.09\pm0.12) \times ~10^{-4}~$, where the first uncertainty is
 statistical and the second systematic. This result is consistent with the previously published BESIII measurement but with an improvement in accuracy
 by a factor of 1.4.

\begin{acknowledgments}
The BESIII collaboration thanks the staff of BEPCII and the IHEP computing center for their strong support. This work is supported in part by National Key Basic Research Program of China under Contract No. 2015CB856700; National Natural Science Foundation of China (NSFC) under Contracts Nos. 11625523, 11635010, 11735014, 11822506, 11835012; the Chinese Academy of Sciences (CAS) Large-Scale Scientific Facility Program; Joint Large-Scale Scientific Facility Funds of the NSFC and CAS under Contracts Nos. U1532257, U1532258, U1732263, U1832207; CAS Key Research Program of Frontier Sciences under Contracts Nos. QYZDJ-SSW-SLH003, QYZDJ-SSW-SLH040; 100 Talents Program of CAS; INPAC and Shanghai Key Laboratory for Particle Physics and Cosmology; ERC under Contract No. 758462; German Research Foundation DFG under Contracts Nos. Collaborative Research Center CRC 1044, FOR 2359; Istituto Nazionale di Fisica Nucleare, Italy; Koninklijke Nederlandse Akademie van Wetenschappen (KNAW) under Contract No. 530-4CDP03; Ministry of Development of Turkey under Contract No. DPT2006K-120470; National Science and Technology fund; STFC (United Kingdom); The Knut and Alice Wallenberg Foundation (Sweden) under Contract No. 2016.0157; The Royal Society, UK under Contracts Nos. DH140054, DH160214; The Swedish Research Council; U. S. Department of Energy under Contracts Nos. DE-FG02-05ER41374, DE-SC-0010118, DE-SC-0012069; University of Groningen (RuG) and the Helmholtzzentrum fuer Schwerionenforschung GmbH (GSI), Darmstadt.
\end{acknowledgments}

\end{document}